# Specific Heat of $(Ca_{1-x}Sr_x)_3Ru_2O_7$ Single Crystals

V. Varadarajan, S, Chikara, V. Durairaj, X.N. Lin, G. Cao, and J.W. Brill*

Department of Physics and Astronomy, University of Kentucky

Lexington, KY 40506-0055, USA

ABSTRACT

We have measured the specific heat of crystals of $(Ca_{1-x}Sr_x)_3Ru_2O_7$ using ac- and relaxation-time calorimetry. Special emphasis was placed on the characterization of the Néel ($T_N$=56 K) and structural ($T_c$ = 48 K) phase transitions in the pure, x=0 material. While the latter is believed to be first order, detailed measurements under different experimental conditions suggest that all the latent heat (with L ~ 0.3 R) is being captured in a broadened peak in the effective heat capacity. The specific heat has a mean-field-like step at $T_N$, but its magntitude ($\Delta c_P \sim R$) is too large to be associated with a conventional itinerant electron (e.g. spin-density-wave) antiferromagnetic transition, while its entropy is too small to be associated with full ordering of localized spins. The $T_N$ transition broadens with Sr substitution while its magnitude decreases slowly. On the other hand, the entropy change associated with the $T_c$ transition decreases rapidly with Sr substitution and is not observable for our x=0.58 sample.

The Ruddlesden-Popper ruthenates, $(Ca_{1-x}Sr_x)_{n+1}Ru_nO_{3n+1}$, with $n$ neighboring layers of corner-sharing, distorted $RuO_6$ octahedra separated by alkaline earth layers, exhibit a large variety of magnetic and superconducting states, reflecting the competition between spin, orbital, and charge ordering on the $Ru^{4+}$ S=1 ions [1-3]. In general, substitution of $Sr^{2+}$ with the smaller $Ca^{2+}$ ions tends to increase distortion and tilting of the octahedral with a resulting decrease in valence band width and tendency toward antiferromagnetic ordering [3-6]. The interactions are strong functions not only of the cation, however, but also of the number of layers $n$, even though the basal plane structures for different $n$ are very similar [3]. For example, while single layer $Sr_2RuO_4$ has a superconducting ground state [7], triple layer $Sr_4Ru_3O_{10}$ has competing (interlayer) ferromagnetic and (intralayer) antiferromagnetic interactions [8] which may give rise to quantum critical behavior [9].

The spin-structure of the double-layer, $n$ =2, compounds have been especially unusual. In the pure strontium (i.e. x=1) salt, there is no resulting long range spin order at ambient pressure, but longitudinal strain can induce a ferromagnetic state [5,10]. In the pure calcium (x=0) salt, the spins order ferromagnetically within the bilayers, but the interlayer (c-axis) interactions are predominantly antiferromagnetic, leading to a Néel transition at $T_N$ = 56 K with spin polarization along the $b$-axis [6,11-13]. With further cooling, the spin polarization rotates within the plane toward the $a$-axis [13,14]. There is a second, more unusual, phase transition at $T_c$ = 48 K; here all components of the susceptibility drop abruptly upon cooling through $T_c$ [11,14], which has been suggested to be due to formation of a charge-density-wave [15], while the lattice contracts along $c$ by ~ 0.1 % [13,16]. Crystals grown by a flux method become nonmetallic below $T_C$ [11], but for those grown in a floating zone furnace only the c-axis resistivity becomes activated, so the transition appears to correspond to a change from three-dimensional to two-dimensional transport [16,17]. It is not yet clear which type of growth yields more defect-free (e.g. oxygen-stoichiometric) crystals. Finally, while the $T_c$ transition, on the basis of the abruptness of the structural, magnetic, and resistance changes, appears to be first order, no thermal hysteresis has been observed in any measurement [11-17], so the free energy change between the two states is presumably very small.

In this paper, we report on the specific heat of flux grown pure and Sr-substituted crystals of $Ca_3Ru_2O_7$. Previous measurements, with relaxation-time calorimeters, could resolve the $T_N$ transition in the pure material but the anomaly was too small for quantitative measurements [14,17]. On the other hand, relaxation-time calorimetry is not necessarily a reliable method for measuring latent heats at first order transitions, although all or part of the latent heat (L) may appear in a peak in the measured specific heat. Indeed, very sharp peaks in the effective specific heat were observed [14,17] and the latent heat estimated as L ~ 130 J/mol (formula unit) [14], corresponding to a small entropy change of ~ 0.32 R (where R = 8.314 J mol$^{-1}$ K$^{-1}$ is the gas constant.)

In this work, we use both relaxation-time calorimetry [18] and ac-calorimetry [19-21] to examine the specific heat anomalies at the phase transitions. While ac-calorimetry may also not be a reliable method for obtaining latent heats [19,22], we have compared the specific heat peaks obtained under different experimental conditions and for different samples to conclude that most of the latent heat is in fact appearing in this peak, in agreement with the conclusion of Reference [14]. In addition, we have quantitatively measured the mean-field-like step at $T_N$ for the first time; its size is intermediate between what would be expected for an antiferromagnetic transition of intinerant (e.g. a spin-density-wave) and localized spins. We also discuss the evolution of these anomalies with Sr-substitution.

Single crystals of $(Ca_{1-x}Sr_x)_3Ru_2O_7$ were grown using flux techniques [4]. All single crystals were grown in Pt crucibles from off-stoichiometric quantities of $RuO_2$, $CaCO_3$ ($SrCO_3$) and $CaCl_2$ ($SrCl_2$) mixtures with $CaCl_2$ ($SrCl_2$) being self flux. The mixtures were first heated to 1480 °C in a Pt crucible covered by a Pt cover, soaked for 25 hours, slowly cooled at 2-3 °C/hour to 1380 °C, and finally cooled to room temperature at 100 °C/hour. The starting Ca:Ru ratio and the thermal treatments are critical and subtle for the formation of perovskite crystals as nucleation of its sister compounds (Ca,Sr) $RuO_3$ and $(Ca,Sr)_2RuO_4$ are also energetically favorable. By carefully changing the ratio and thermal treatments, we have successfully grown crystals of $Ca_{n+1}Ru_nO_{3n+1}$ and $Sr_{n+1}Ru_nO_{3n+1}$ with n=1, 2, 3 and $\infty$ [3,11,20]. All crystals studied were characterized by single crystal or powder x-ray diffraction, EDS and TEM, indicating good crystal quality with no impurities or intergrowths or significant clustering of strontium ions. Magnetic and transport properties were measured using a Quantum Design MPMS XL 7T magnetometer. (Since the shape of all crystals studied in this work is essentially cubic, the demagnetizing factor for all samples and all three principal crystallographic axes is expected to be the same, and is not included in our normalization.)

Results of *ab*-plane magnetization measurements, measured with a small field (B=0.5 T) for the same samples measured with ac-calorimetry, are shown in Figure 1. For the pure x=0 sample, the $T_N$ = 56 K transition appears as a peak in $M_b$ and cusp in $M_a$, while $M_a$ falls at the $T_c$=48 K transition, as previously observed [11]. "Light" (x=0.15) Sr substitution increases $T_N$ and decreases $T_c$, decreases the size of the magnetic anomalies (with $M_b$ developing a minimum between the two transitions), and makes the overall magnetic susceptibility more isotropic. A transition to a nonmetallic state is also observed in the resistance (not shown) between 40 K and 50 K.  For the x=0.58 sample, the susceptibility is almost isotropic in the ab-plane (suggesting that the $RuO_6$ octahedra are almost "untilted", as in the pure Sr compound) with peaks in M at $T_p \sim 50$ K followed by weak minima (T~ 40 K) and relatively high susceptibilities at lower temperatures. Only small anomalies are observed in the resistance at these temperatures, and the shape of the magnetic anomalies [4] suggests that spin ordering is complex but incomplete.  A more complete discussion of the substitution-dependence of the resistance and magnetization, including their field dependences, will be presented elsewhere.

We used ac-calorimetry [19-21] to measure the specific heat of flux grown pure and Sr-substituted single crystals. For the pure materials, measurements were made on two crystals of masses 0.36 and 0.40 mg.  For the Sr-substituted samples, the masses were 0.43 mg (x=0.15) and 0.54 mg (x=0.58). Each sample was attached with silver paint to a flattened, 25 μm-diameter type-E thermocouple junction. Light from a quartz halogen lamp, chopped at frequency ω, was used to heat the sample. The magnitude of the resulting temperature oscillation, $\Delta T(\omega)$, is inversely proportional to the heat capacity of sample (and addenda) if the chopping frequency satisfies $\omega\tau_{int} << 1 << \omega\tau_{ext}$, where $\tau_{int}$ and $\tau_{ext}$ are the internal and external thermal time constants [19].  The typical chopping frequency was $\omega/2\pi \sim 6$ Hz (with resulting $\Delta T(\omega) \sim 3$ mK), although, as discussed below, measurements were taken at several frequencies near the phase transitions, where, the temperature drift rate, |dT/dt|, was kept less than 0.1 K/min.

Because our ac-calorimetric measurements only yield relative data [19], the results for the pure (x=0) samples were normalized (at T=130 K) and corrected for addenda contributions (~ 10% at low temperatures) to the data taken on a pure 0.78 mg sample with a Quantum Design PPMS relaxation-time calorimeter [18], as shown in Figure 2. Note that we were not able to obtain quantitative results on this sample with ac-calorimetry because it had a relatively long internal time constant, presumably because of an interlayer crack. On the other hand, PPMS results on the smaller "ac" samples were not also quantitative; in fact, the 0.76 mg sample heat capacity was only ~ 10% of the PPMS platform heat capacity for temperatures > 20 K, so our normalizing data should only be considered approximate. We note, however, that our PPMS results are within 10% of the results of Yoshida *et al* [17], except at the lowest temperatures. A fit of our low temperature PPMS results (Figure 2b) to $c_P = \gamma T + \beta T^3$ for 10 K < T < 40 K gives values of $\gamma/R \sim 1.3 \times 10^{-3}$ K$^{-1}$ and $\beta/R \sim 4.8 \times 10^{-5}$ K$^{-3}$ (corresponding to a Debye temperature of $\Theta = 388$ K), intermediate between the values of Yoshida *et al* [17] and McCall *et al* [14].

One consequence, however, of normalizing our ac results to the PPMS results at high temperature is that, as seen in Figure 2b (and also Figure 4), the low-temperature ac results differ from the PPMS results by ~ 10%, reflecting both the uncertainty in the PPMS results and marginal sample time constants at some temperatures. For the strontium substituted samples, which were not large enough for quantitative PPMS measurements, we normalized the data by assuming that the force constants are roughly equal in all materials so that the low temperature $\beta \sim$ (molecular weight)$^{3/2}$ (approximately consistent with the results of Cao *et al* [4]), as also shown in Figure 2b, with the resulting overall temperature dependences shown in Figure 2a. Hence, for all samples, the normalization of the data should only be considered approximate (roughly ±20 %).

PPMS and ac results near the lower transition for the pure samples are shown in Figure 3. As mentioned above, there are difficulties in measuring latent heat with both ac and relaxation-time calorimetry. Certainly, if free energy barriers are so large so that thermal hysteresis remains significant even for our slow drift speeds, the latent heat will not be observed at all with ac-calorimetry, and will be improperly characterized with the PPMS measurement [18]. More subtly, if the latent heat has a time constant much greater than $\tau_{int}$ (away from the transition), the resulting measured latent heat can be frequency dependent [19,22]. However, if the hysteresis is negligible (i.e. less than $\Delta T(\omega)$ or the step size, for ac and PPMS calorimetry respectively) and the transition "fast", the latent heat contributes to the effective measured specific heat as $c_P(effective) = c_P + dL/dT$ [19], where we assume that the latent heat is distributed over a temperature interval due to sample inhomogeneity (or experimental temperature gradients in the sample). In fact, no significant hysteresis in the anomalies was observed ($\Delta T < 0.1$ K).

For the PPMS sample, measurements were made with several different size temperature steps, with very little change in the thermal anomaly, as shown in Figure 3 for step sizes of 50 and 200 mK, suggesting that, unless the hysterisis is much wider, most of the latent heat has been captured. Similarly, for the 0.36 mg sample, we measured the specific heat for a variety of chopping frequencies (with $\Delta T(\omega) \sim 1/\omega$), again with negligible difference in results, as shown in the Figure 3, also indicating that the transition is not "sluggish". For all samples, the anomaly is very sharp, $\Delta T \sim 0.25$ K, as observed by others [14,17]. The measured transition entropy, $\Delta S \equiv \int dT \, \Delta c_P / T$, is somewhat sample dependent but reasonably well-defined, with an average value $\Delta S = (0.31 \pm 0.08)$ R, very close to the value measured by McCall, *et al* [14]. Again, the robustness of this value suggests that at least most of the latent heat is being captured in the specific heat measurements. If this transition is due to formation of a charge-density-wave, as suggested by Baumberger, *et al* [15], then the expected entropy change $\Delta S \sim \Delta\gamma \cdot T_c$ (assuming that fluctuation effects are negligible [19,21,23]), giving $\Delta\gamma / R \sim 6 \times 10^{-3}$ $K^{-1}$, i.e. the change in $\gamma$ at the transition is significantly larger than its low temperature value, consistent with the low conductivity and/or lower dimensionality of the low temperature state.

Results for the three pure samples at $T_N$ transition are shown in Figure 4; the anomalies are very similar. While our $c_P$ anomalies are about twice those measured by McCall, *et al* [14], a simple estimate of the entropy change using the baseline shown, $\Delta S \sim 0.036R$, is very close to that of Reference [14]. Of course, the proximity of the large $T_c$ anomaly makes reliably estimating an appropriate baseline difficult. It is clear, however, that $\Delta S \ll 2R \ln(3)$, the value expected for complete ordering of localized S=1 spins. On the other hand, for antiferromagnetic ordering of itinerant spins, e.g. in a spin-density-wave, we again expect $\Delta S \sim \Delta\gamma \cdot T_N$ [24,25]. Formation of a spin-density-wave is also consistent with the mean-field-like step in the specific heat, for which we expect $\Delta c_P \sim 1.43 \, \Delta\gamma \cdot T_N$. Then our measured anomaly $\Delta c_P \sim 1.0$ R gives $\Delta\gamma / R \sim 0.012 \, K^{-1}$. This value, greater than not only the low-temperature value of $\gamma$ but also the possible change at $T_c$, is much larger than expected, since the changes in conductivities at $T_N$ are small (and, in fact, the conductivity increases with cooling through $T_N$). While the anomaly may be enhanced by low-dimensional fluctuations (effectively suppressing $T_N$ below the mean-field transition temperature [21,23-25]), this is unlikely to be a large effect here. Hence the specific heat anomaly at $T_N$ is inconsistent with both conventional itinerant and localized pictures of antiferromagnetic ordering, suggesting a more complex spin ordering. (We note that a somewhat smaller but similar shape anomaly was observed at the Curie point in triple-layer $Sr_4Ru_3O_{10}$ [20].)

The effects of strontium substitution on the thermal anomalies is shown in Figure 5. For the x=0.15 sample, the $T_N$ and $T_c$ anomalies separate to ~ 60 K and 45 K, consistent with the magnetization data. The $T_c$ anomaly is approximately symmetric, suggestive of a broadened first order transition; our estimate of the entropy, with the baseline shown, is $\Delta S$ ~0.08R, much smaller than for the pure sample, suggesting a smaller structural change and, if electronic in origin, smaller change in $\gamma$. In contrast, the specific heat above $T_N$ extrapolates to a lower value than that measured below $T_N$, as shown in the figure, so the anomaly appears to be "broadened mean-field", with $\Delta c_P$ ~ 0.7 R, slightly smaller than for the pure sample. For the x=0.58 sample, we observe only a single, "broadened mean-field" anomaly at $T_N$ ~ 53 K, with $\Delta c_P$ ~ 0.4 R. The decrease in magnitude of the $c_P$ anomalies at $T_N$ with Sr-substitution parallels the decrease in size of the magnetic anomalies, suggesting that Sr-substitution decreases either the magnitude or the average density of ordering moments. The absence of a second anomaly for the x=0.58 sample suggests that the small resistance anomalies observed for this sample below 50 K are not due to thermodynamic transitions, but to changes in scattering as the spin state evolves.

In conclusion, we have reported on detailed studies of the specific heat of $(Ca_{1-x}Sr_x)_3Ru_2O_7$ crystals.  For pure (x=0) samples, we find that the large peak in effective specific heat at the $T_c$ = 48 K structural transition is independent of measurement times/frequencies, suggesting a latent heat L ~ 0.31 R, which decreases rapidly in magnitude with Sr substitution.  The Néel transition anomaly is roughly mean-field in shape, but its magnitude ($\Delta c_P$ ~ R) is too large for a conventional itinerant electron spin-density-wave transition but too small to be associated with complete spin-ordering of localized electrons.  The magnitude of the anomaly decreases more slowly as Sr substitution decreases the spin-order.

**Acknowledgements**

    This research was supported by U.S. National Science Foundation grants DMR-0400938 and DMR-0552267.

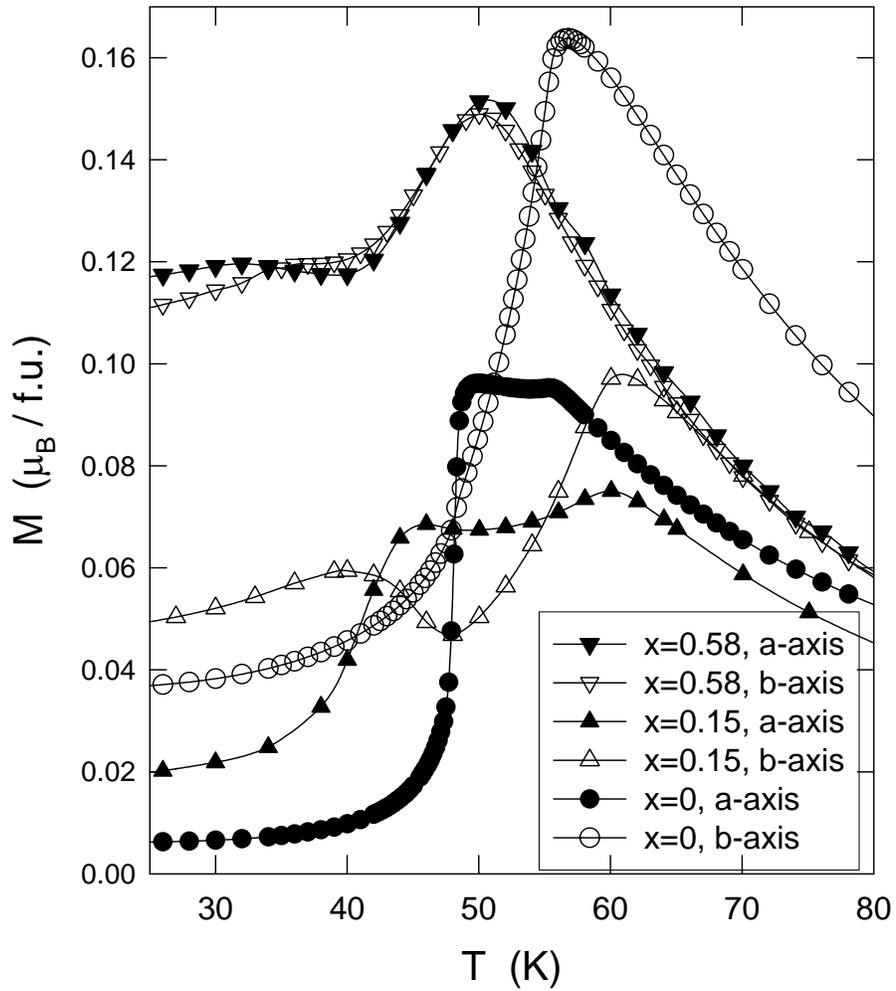

*Figure 1.* Temperature dependence of the magnetic moments (per formula unit) for the pure and strontium substituted samples near the phase transitions with fields (0.5T) applied along the *a*-axis and *b*-axis.

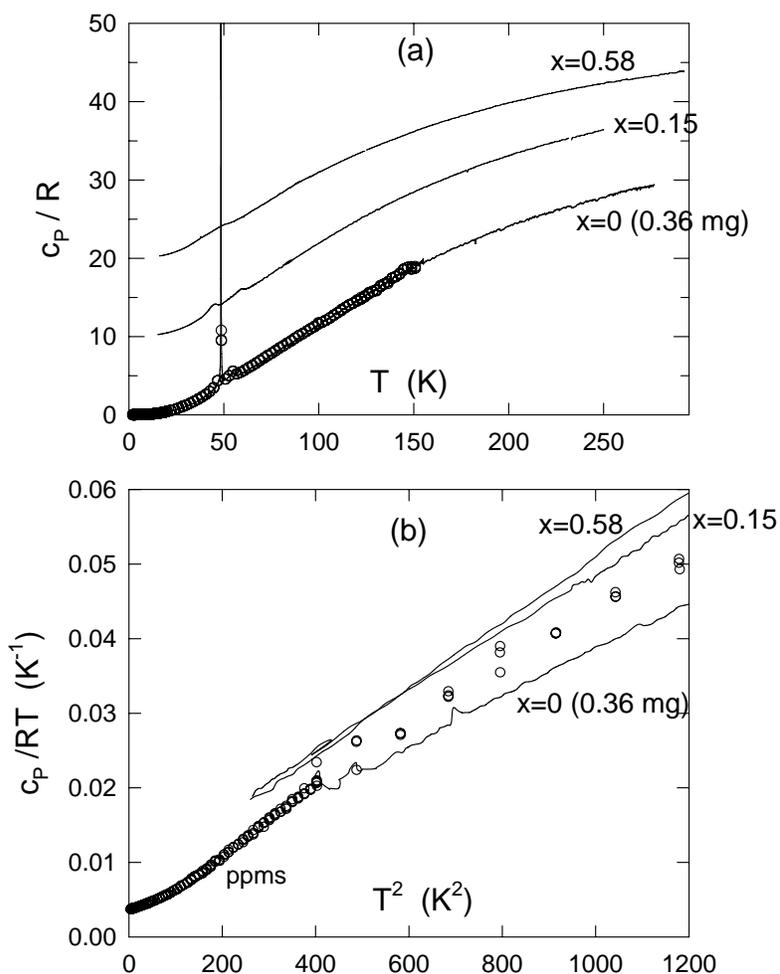

*Figure 2*. (a) The overall temperature dependence of specific heat, normalized to the gas constant R = 8.31 J·mol⁻¹·K⁻¹, of $(Ca_{1-x}Sr_x)_3Ru_2O_7$ crystals measured with ac-calorimetry (solid curves);  data for the x=0.15 and x=0.58 samples are vertically offset by 10 and 20 units, respectively.  The open symbols show the PPMS results for the pure sample.  (b) Enlargement of the low temperature specific heat, plotted as $c_P/RT$ vs. $T^2$, for the same samples.

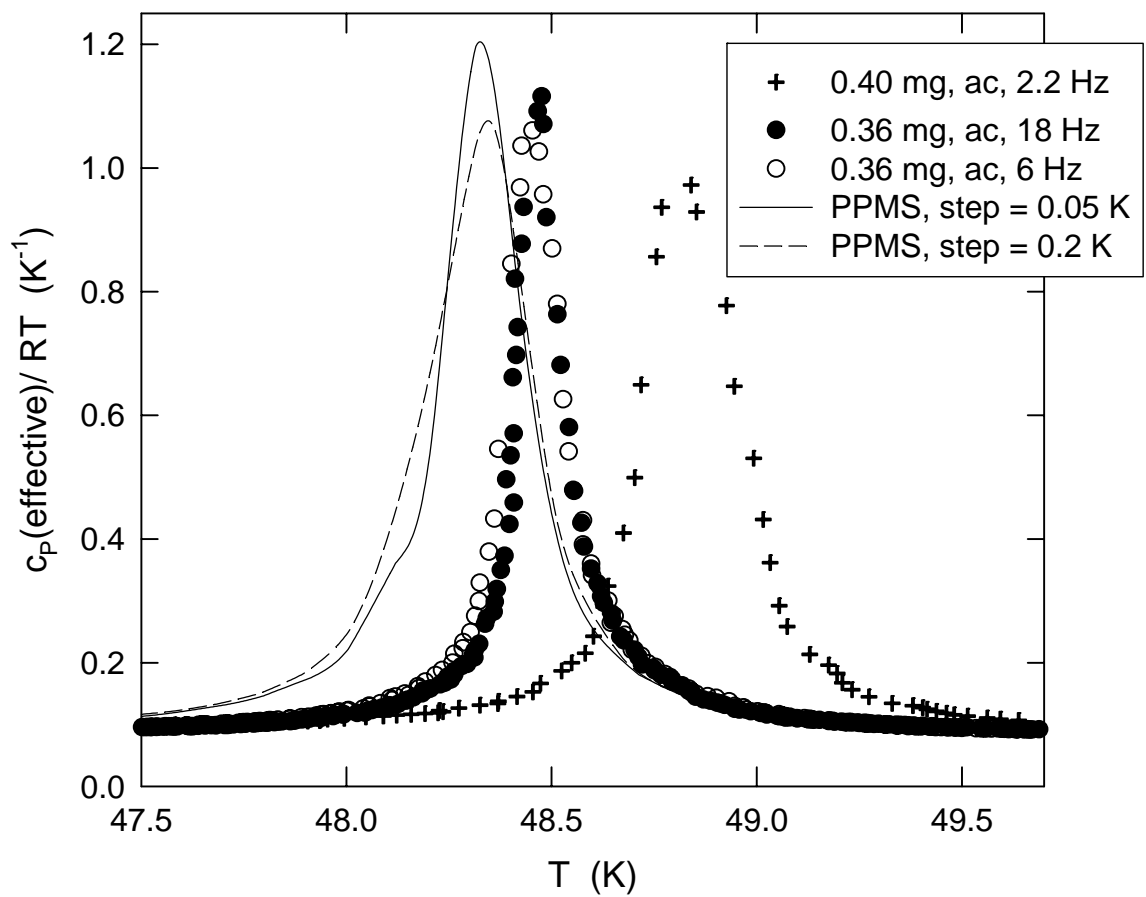

*Figure 3.* Effective specific heat (including distributed latent heat) for the pure samples near the T$_c$ structural transition, measured with different frequencies (ac-calorimetry) or temperature steps (PPMS measurements), as indicated.

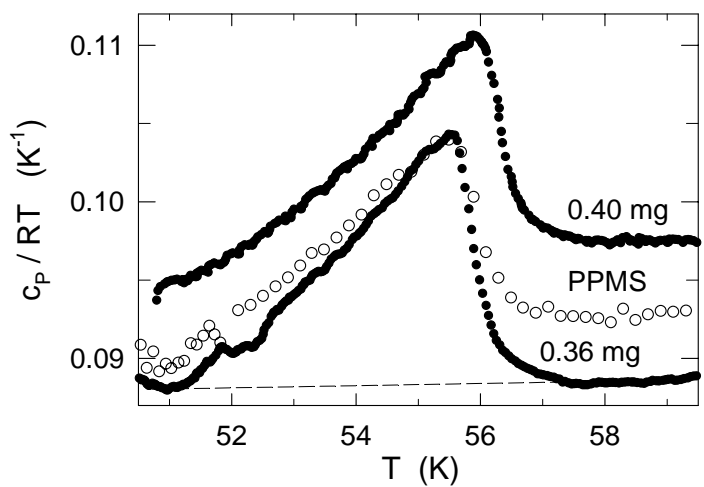

*Figure 4.* Specific heats of the pure samples near their Neel transitions. The dashed line

for the 0.36 mg sample shows the background used to estimate the change in entropy.

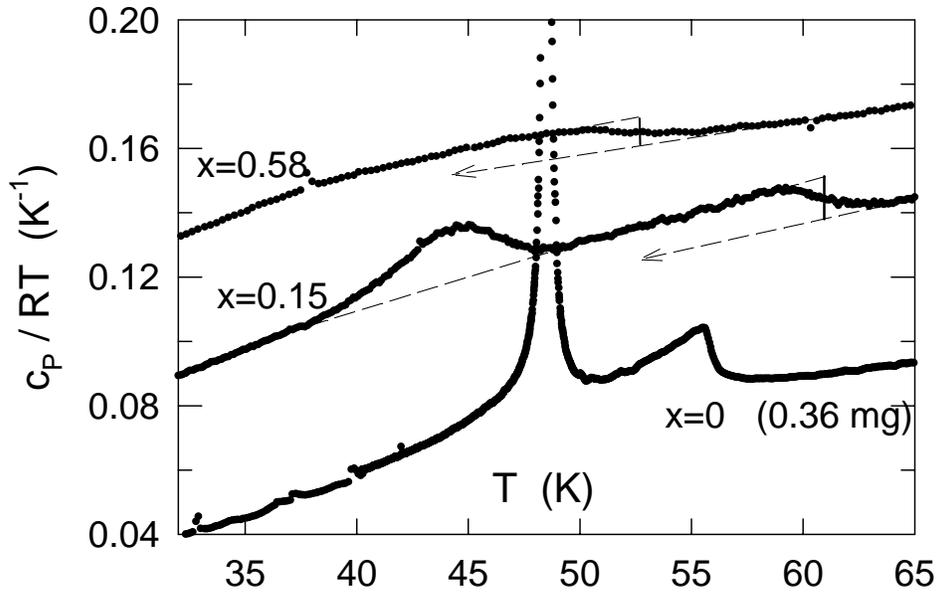

*Figure 5.* Specific heat, measured with ac-calorimetry of pure and Sr-substituted samples, near the phase transitions. Data for the x=0.15 and x=0.58 samples are vertically offset by 0.4 K$^{-1}$ and 0.8 K$^{-1}$, respectively. The dashed lines show the extrapolated temperature dependences and background used to estimate the changes in specific heats (shown by the heavy vertical lines) and entropy, respectively, for the substituted samples, as discussed in text.